\documentclass[12pt]{article}
\usepackage{amsmath}
\usepackage{graphicx}
\usepackage{enumerate}
\usepackage{natbib}
\usepackage[pagebackref,breaklinks,colorlinks,citecolor=blue]{hyperref}
\usepackage{hypernat}
\usepackage{url} % not crucial - just used below for the URL 
\usepackage{float}
\usepackage{adjustbox}
\usepackage{mathrsfs}
\newcommand{\blind}{1}
\DeclareMathOperator*{\argmin}{arg\,min}

\usepackage{xcolor}

\usepackage{amsfonts}
\usepackage{caption}

\begin{document}

\bibliographystyle{natbib}
\captionsetup[figure]{font=footnotesize}

\def\spacingset#1{\renewcommand{\baselinestretch}%
{#1}\small\normalsize} \spacingset{1}

%%%%%%%%%%%%%%%%%%%%%%%%%%%%%%%%%%%%%%%%%%%%%%%%%%%%%%%%%%%%%%%%%%%%%%%%%%%%%%

\begin{center}
\begin{Large}{\bf Elastic Shape Analysis of Movement Data\\}
\vspace{.25in}
\end{Large}

\if1\blind
{
J.E. Borgert$^{1,2}$, Jan Hannig$^{1}$, J.D. Tucker$^{3}$, Liubov Arbeeva$^{2}$, Ashley N. Buck$^{2,4,5}$, \\Yvonne M. Golightly$^{2,6,7}$, Stephen P. Messier$^{8}$, \\Amanda E. Nelson$^{2,7,9}$, J.S. Marron$^{1}$\\[.15in]
$^{1}$Department of Statistics $\&$ Operations Research, University of North Carolina, Chapel Hill, NC \\
$^{2}$Thurston Arthritis Research Center, University of North Carolina, Chapel Hill, NC\\
$^{3}$Statistical Sciences, Sandia National Laboratories, Albuquerque, NM\\
$^{4}$Human Movement Science Curriculum, University of North Carolina, Chapel Hill, NC\\
$^{5}$Department of Exercise $\&$ Sport Science, University of North Carolina, Chapel Hill, NC\\
$^{6}$College of Allied Health Professions, University of Nebraska Medical Center, Omaha, NE\\
$^{7}$Department of Epidemiology, Gillings School of Global Public Health, University of North Carolina, Chapel Hill, NC\\
$^{8}$Department of Health $\&$ Exercise Science, Wake Forest University, Winston-Salem, NC
$^{9}$Department of Medicine, University of North Carolina, Chapel Hill, NC\\

} \fi
\end{center}

\bigskip
\begin{abstract}
Osteoarthritis (OA) is a highly prevalent degenerative joint disease, and the knee is the most commonly affected joint. Biomechanical factors, particularly forces exerted during walking, are often measured in modern studies of knee joint injury and OA, and understanding the relationship among biomechanics, clinical profiles, and OA has high clinical relevance. Biomechanical forces are typically represented as curves over time, but a standard practice  in biomechanics research is to summarize these curves by a small number of discrete values (or \textit{landmarks}). The objective of this work is to demonstrate the added value of analyzing full movement curves over conventional discrete summaries. We developed a shape-based representation of variation in full biomechanical curve  data from the Intensive Diet and Exercise for Arthritis (IDEA) study \citep{messier.2009.bmc,messier.2013.jama}, and demonstrated through nested model comparisons that our approach, compared to conventional discrete summaries, yields stronger associations with OA severity and OA-related clinical traits.  Notably, our work is among the first to quantitatively evaluate the added value of analyzing full movement curves over conventional discrete summaries.
\end{abstract}

\noindent%
{\it Keywords:}  Functional data analysis, Biomechanical data, Nonparametric methods, Shape statistics
\vfill

\newpage
\spacingset{1.7} % DON'T change the spacing!
\section{Introduction}
\label{sec:intro}
Osteoarthritis (OA) is a prevalent degenerative joint disease characterized by cartilage loss, bone and soft tissue changes, joint pain, and diminished function. In the United States, OA affects at least $19\%$ of adults aged 45 years and older, with the knee being the most commonly affected joint, accounting for more than $80\%$ of the total burden of the disease \citep{dillon.2006.jorh,jordan.2007.jorh,lawrence.2008.arthritis,vos.2012.lancet}. Knee OA is characterized by both the severity of radiographically-assessed damage and clinical symptoms, such as knee pain and function. Previous research, such as in \citet{zhang.2010.epidemiology}, indicated that risk factors and disease progression may vary by clinical phenotype. Additionally, important work like \citet{felson.2013.osteoarthritis} and \citet{guilak.2011.biomechanical} identified biomechanical factors in the etiology and pathogenesis of knee OA. 

Biomechanical variables, particularly forces exerted during walking, are often measured in modern studies of knee joint injury and OA, and understanding the relationship among biomechanics, clinical profiles, and OA has high clinical relevance.  Biomechanical forces are typically represented as curves over time, but a standard practice in biomechanics research is to summarize these curves by a small number of discrete values. Such discrete summaries are called \textit{landmarks} in the shape statistics terminology of \citet{dryden.2016.shapestat}. Analyses based on conventional discrete summaries, such as those by \citet{sims.2009.jwomen} and \citet{astephen.2008.jortho}, have identified differences between groups (e.g., sex differences) and discovered variations in gait patterns associated with knee OA-related outcomes. Recent work by \citet{buck.2024.biomechanical} evaluated the ability of various clinical traits and conventional discrete summaries of gait forces to predict early symptomatic knee OA.  
While simplifying the statistical methods required for analysis, relying on conventional discrete summaries risks overlooking information encoded in the complete range and patterns of movement data. Research by \citet{ieee.gait}, \citet{davis.2019.sagittal}, \citet{costello.2021.oac}, \citet{bjornsen.2024.arthritis}, and others indicate the value of analyzing full movement curves. However, these studies are limited and did not formally compare analyses based on conventional discrete summaries. For harmonic analyses of full movement curves, see \citet{trentadue.2024.fourier} and references therein.

The primary goal of this work is to demonstrate the added value of analyzing full movement curves over conventional discrete summaries.  Our analysis of full movement curves follows an Object Oriented Data Analysis (OODA) approach. OODA, described in \citet{ooda}, is a framework for analyzing complex data that emphasizes the careful selection of data objects for a given scientific question and the utilization of methods intrinsic to the data object space. This approach facilitates the consideration of full movement data curves as complex data objects in high-dimensional space. Furthermore, the richer information within these curves allows for potentially many different choices of both the \textit{data object} of interest and the appropriate methodology for analysis.

We developed a shape-based representation of variation in  full biomechanical curves using data from the Intensive Diet and Exercise for Arthritis (IDEA) study \citep{messier.2009.bmc,messier.2013.jama}, and demonstrated through easily interpretable nested model comparisons that our approach, compared to conventional discrete summaries, yields stronger associations with OA severity and OA-related clinical traits. Notably, our work is among the first to quantitatively evaluate the added value of analyzing full movement curves over conventional discrete summaries in OA research.

\section{Curves as Data Objects}\label{sec:data}
During gait data collection in the IDEA study, participants wore laboratory-provided cushioned shoes and walked at their preferred speed on a 22.5m walkway. Kinetic data, including Ground Reaction Force (GRF), were collected using an Advanced Medical Technologies, Inc. model OR-6-5-1 force plate (480 Hz) and filtered using a $4^{\text{th}}$ order low-pass Butterworth filter with a cutoff frequency of 6 Hz. The GRF consists of three components: vertical GRF (vGRF), representing the force exerted downwards; anterior-posterior GRF (apGRF), the propulsive or braking force in the direction of walking; and medial-lateral GRF (mlGRF) in the third orthogonal direction. When representing GRF as a function of time, the horizontal axis reflects the percentage of \textit{stance} (time interval of foot contact with the ground) over a gait cycle, and the vertical axis corresponds to body weight-normalized force values (measured in Newtons).

A walking trial was recorded as a successful observation when the participant's entire foot maintained contact with the force plate throughout the stance phase, and participants maintained their preferred walking speed within $\pm 3.5\%$. Walking speed was defined as the mean speed at which participants walked a 10m walkway at a self-selected pace over six practice trials. A photocell system registered speed and provided participants with real-time visual feedback for maintaining their walking speed during formal trials.

We studied the part of the IDEA data that contained measurements of each GRF component taken at a constant sampling rate over the duration of each step. For each IDEA participant, three trials per limb were collected. For some participants, one or more trials were missing from the data. In those cases, we used as many trials as were given in the data and did not impute. We considered the collection of all trials of both limbs from all participants.

\begin{figure}[H]
	\centering
	\includegraphics[trim=0cm 8cm 2cm 8cm,clip=true,width=0.45\textwidth]{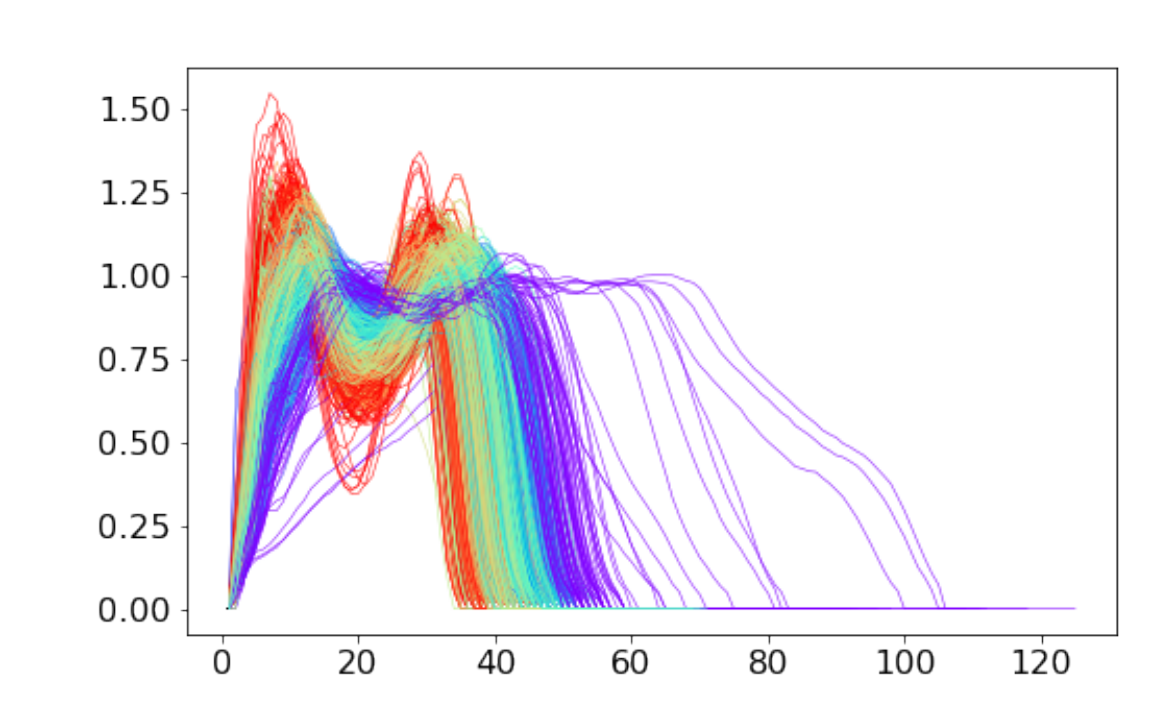}
        \includegraphics[trim=0cm 8cm 2cm 8cm,clip=true,width=0.45\textwidth]{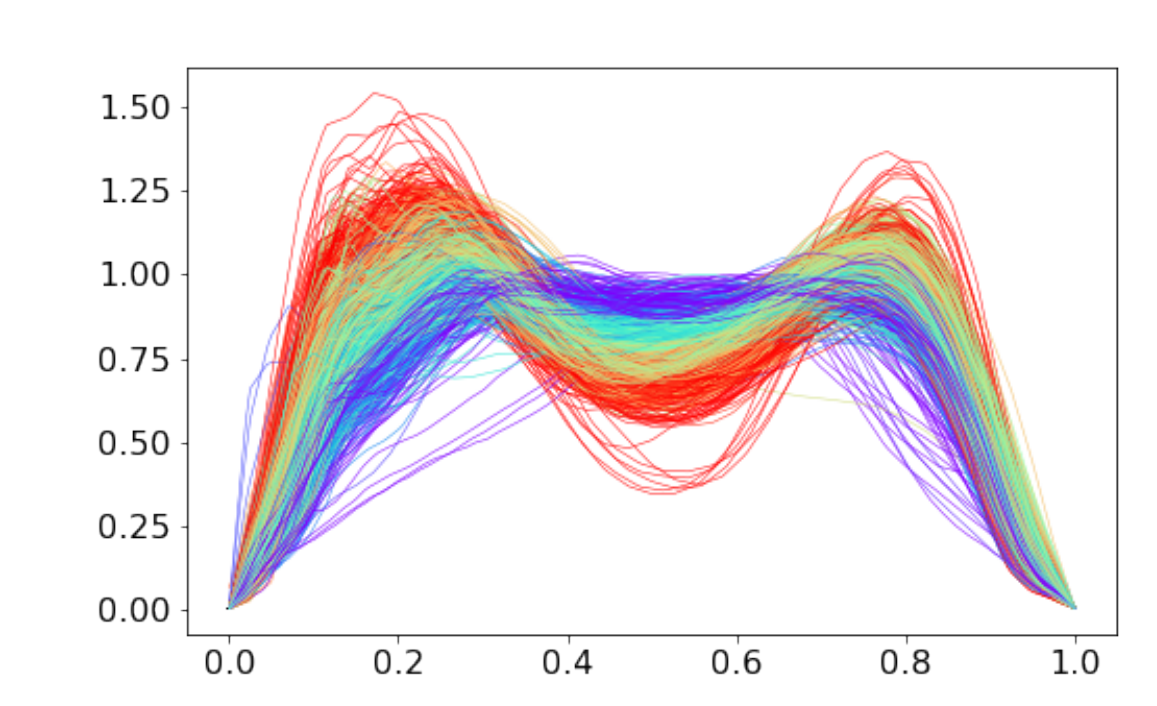}
        \includegraphics[trim=0cm 8cm 2cm 8cm,clip=true,width=0.45\textwidth]{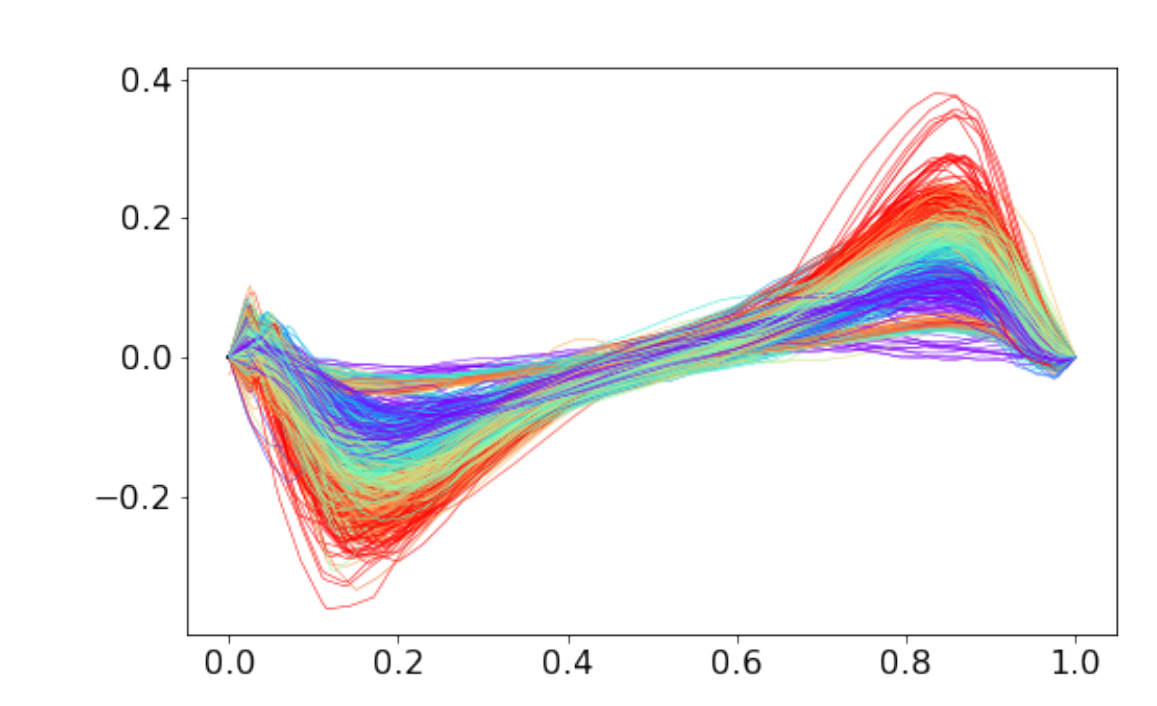}
	\includegraphics[trim=0cm 8cm 2cm 8cm,clip=true,width=0.45\textwidth]{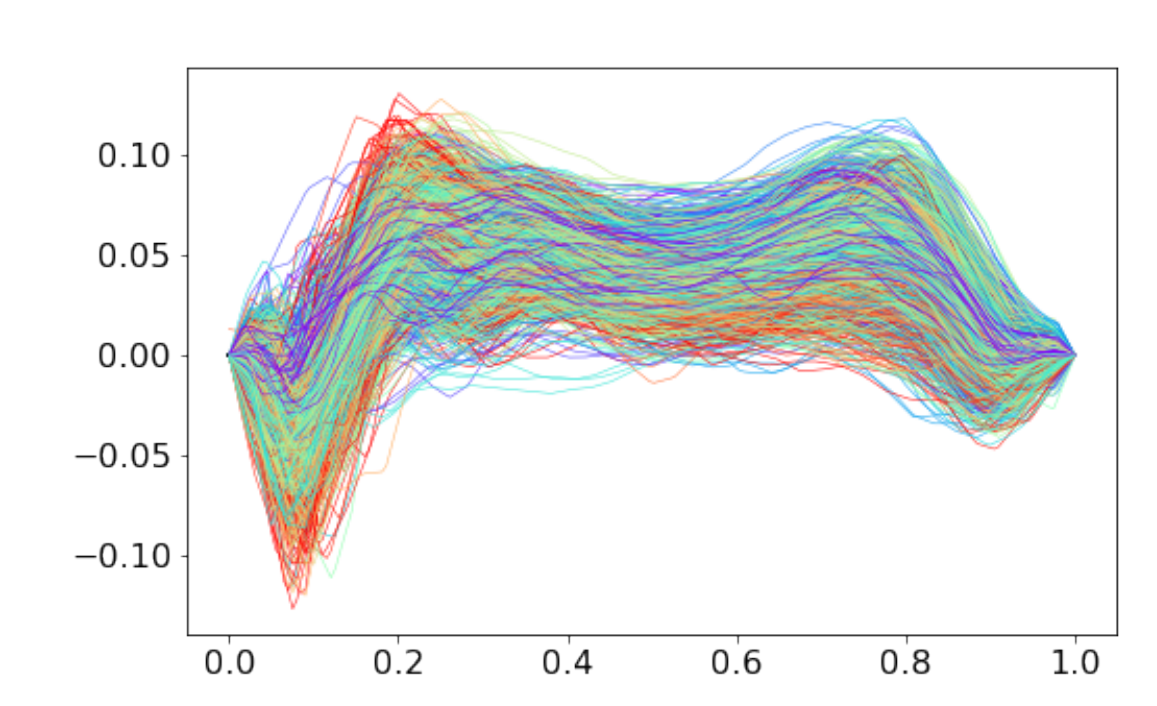}
	\caption{Top left panel: Raw vertical ground reaction force (vGRF) curves before processing, colored by walking speed. Top right: The same curves after re-scaling and interpolation of the time axis . The common rainbow color palette descends from fastest walking speed (red) to slowest (purple). Bottom row panels show apGRF curves (left) and mlGRF curves (right) after the same processing of the time axis. Note that the vertical axis (measured in N/kg, i.e. percentage of body weight) of each panel is scaled to the data it displays.}
	\label{fig:curvepreprocess}
\end{figure}

The complete set of curves for each GRF component contained 2,686 curves from 454 participants. The top left panel of Figure~\ref{fig:curvepreprocess} shows the collection of raw data vGRF curves colored by walking speed. There, the rainbow descends from fastest walking speed (red) to slowest (purple). Notice that many of those curves have consecutive starting and trailing zeros, which are outside the stance phase (i.e., after the foot has left contact with the force plate, but force data were still being collected) and hence do not correspond to a meaningful part of the measurements. To account for these spurious measurements, we take the beginning of the force curve as the zero value immediately preceding the first nonzero value, and similarly, the end of the curve as the zero value immediately following the last nonzero value. Using the relevant segment of each curve, we re-scaled the horizontal axis to the unit interval [0,1] in order to establish a common time axis across all curves. Additionally, we applied linear interpolation to the force values, aligning them to an evenly spaced grid. Time normalization to the stance phase $(0–100\%)$ is a standard approach in biomechanics for analyzing GRF and other kinetic variables. The most commonly used method for time normalization is \textit{linear length normalization}, which is a rescaling of the stance phases to a standard interval \citep{helwig.2011.gaitalign}, as was applied in our analysis. While this approach removes explicit information about stance duration, it retains timing differences in key joint-loading events, such as heel-strike and toe-off, which impact cartilage stress and are therefore relevant to understanding OA. In this work, we chose to focus on a careful analysis of shape, and leave analyzing stance duration as a direction for future work. However, stance duration is strongly correlated with walking speed and distance \citep{hebenstreit.2015.subphase}, which were retained as clinical variables in our analysis.

The top right panel of Figure~\ref{fig:curvepreprocess} presents the collection of vGRF curves shown in the first panel, following this processing. Those curves are also color-coded according to the participant's walking speed. Similar results were achieved using the same processing steps for the apGRF and mlGRF curves, and are shown in the bottom left and bottom right panels of Figure~\ref{fig:curvepreprocess}, respectively. For a more detailed view of GRF curve variations across walking speeds, additional plots showing subsets of the curves grouped by deciles of walking speed are provided in Section~{1} of the supplementary material.

\subsection{Conventional discrete summaries}
Most relevant studies of GRF data, such as those in \citet{messier.1992.kneeoa}, \citet{hunt.2006.associations}, \citet{zeni.2009.differences}, and \citet{wiik.2017.abnormal}, rely on discrete summaries (also called landmarks) rather than analyzing the full curves. These discrete summaries typically correspond to critical subphases of stance during gait, such as the first and second vGRF peaks (heel-strike and toe-off) and positive and negative peaks of the apGRF curve (braking and propulsion). In contrast, mlGRF curves are highly variable and therefore more challenging to interpret and summarize \citep{costello.2021.oac}. Consequently, GRF studies primarily analyze only the vGRF and apGRF directions.

Traditionally, the first vGRF peak has been of greater interest, as heel-strike is a significant factor in joint compressive loads and has been consistently linked to knee OA onset and progression. However, recent work by \citet{buck.2024.biomechanical} (and references therein) indicated that the valley (reflecting mid-stance) and the second peak of the vGRF curve are better predictors of knee OA-related symptoms.

In this analysis, we focus on three conventional discrete summaries: (1) the first vGRF peak, (2) the apGRF braking peak, and (3) the apGRF propulsion peak. Throughout the remainder of this paper, we refer to the 3-dimensional vector containing these landmark values as the \textit{conventional discrete summaries}. While other landmark choices could have been analyzed, as discussed in the preceding paragraphs, these three were selected because they are the most commonly studied GRF features in biomechanical research on knee OA.

The first vGRF peak is defined in the references aforementioned as the maximum value within the first $50\%$ (0-$50\%$, heel-strike to midstance) of the stance phase (0-$50\%$) (0-0.5 on the horizontal axis in Figure~\ref{fig:curvepreprocess}). Similarly, the apGRF braking peak is defined as the minimum value over the first $50\%$ of the stance phase, and the apGRF propulsion peak is defined as the maximum value over the second $50\%$ ($50-100\%$, mid-stance to toe-off) of the stance phase (0.5-1.0 on the horizontal axis in Figure~\ref{fig:curvepreprocess}).

\section{Shape-based Functional Data Analysis}\label{sec:methods}

\subsection{Elastic shape analysis}\label{ss:registration}

The goal of our analysis was to characterize patterns of variation in gait using information in the full force curves. Many datasets of curve data have variation that appears to be either vertical or horizontal in nature. This variation is termed \textit{amplitude} and \textit{phase} variation, respectively. In this context, horizontal variation is viewed as a potentially important aspect of gait, while in other contexts it may represent temporal misalignment. The curves shown in Figure~\ref{fig:curvepreprocess} exhibit interesting variation of both types. In particular, there is clear phase variation in the timing of the vGRF and mlGRF peaks, as well as the shift from posterior to anterior force in the apGRF curves. 

\textit{Elastic warping} of the time axis can provide aligned curves that better capture amplitude variation, as studied in Section~{2.1} of \citet{ooda}. This is important because poor alignment of curves due to phase variation can impact statistical methodology, potentially obscuring important geometric structure. As noted by \citet{helwig.2011.gaitalign}, temporal alignment is a critical consideration in gait analysis, with various approaches applied in gait studies. That paper highlights that while linear length normalization, which we applied as a pre-processing step, removes differences in stance duration, it does not account for phase variation in the timing of key events (e.g., peaks and valleys). Other methods discussed by \citet{helwig.2011.gaitalign} require extensive manual tuning and may introduce distortions in curve shape. The results discussed in that paper support the need for an alignment method that captures phase variation while preserving curve shape and minimizes manual tuning.

Elastic warping involves a transformation of the time axis, which is described by a curve that can be usefully thought of as a stretching and compression of the horizontal axis. The functions are aligned by finding the \textit{Karcher mean}~\citep{tucker-wu-srivastava:2013} which produces aligned functions and warping functions and will be defined later on. A warping function $\gamma(x):[0,1] \rightarrow [0,1]$ is strictly increasing, invertible, and \textit{diffeomorphic}, meaning the function and its inverse are smooth. The collection of such warping functions serve as \textit{phase} data objects. 

Aligning points across functions is often referred to as \textit{registration}. Many conventional Functional Data Analysis (FDA) techniques rely on the $\mathbb{L}^2$ norm, which simplifies computations into point-wise evaluations. While point-wise computations involve \textit{vertical registration}, other methods focus on the \textit{shape} of functions. $L^2$-based methods present well-known challenges, as detailed in \cite{marron.2015.statscience}. \textit{Elastic shape analysis}, as proposed in
\citet{srivastava.2011.arxiv, tucker-wu-srivastava:2013}, uses the warp-invariant Fisher-Rao metric to overcome the limitations of conventional $\mathbb{L}^2$-based alignment techniques. This framework dates back to such seminal work as \citet{younes.1998.computable} and is the first to enable fully automatic (meaning that no manual tuning is needed) shape-based registration. For each curve, the elastic shape analysis method computes the warping function needed to align its peaks to a template mean curve, known as the Karcher mean. The Karcher mean is the curve that lies in the ``center" of all the warped curves, meaning it minimizes the total distance (under the Fisher-Rao metric) between itself and all the aligned curves.

The key idea of this method is to define an equivalence relation between curves. Two curves $f_1(x)$ and $f_2(x)$  are called equivalent, $f_1 \sim f_2$, if there exists a warping function $\gamma$ such that $f_1(\gamma(x))=f_1\circ\gamma(x)=f_2(x).$ Then, the set of all warps of a function $f$, given by $$[f] =\{f\circ\gamma: \gamma \in \Gamma\},$$
is an equivalence class and defines the amplitude ( called \textit{shape} in \citet{srivastava.2011.arxiv,wu.2024.test}) of $f$.  

The Fisher-Rao metric defines a proper distance on the set of such equivalence classes. A natural framework for carrying out the computations required for curve alignment is achieved through a \textit{Square Root Velocity Function} (SRVF) representation, which transforms the Fisher-Rao metric into the standard $\mathbb{L}^2$ metric. The Karcher mean equivalence class is defined using this distance, along with the warping functions needed to align individual functions to the Karcher mean template.
The SRVF of a function $f \in \mathcal{F}$ is given by $q(t) = \text{sign}(\dot{f}(t))\sqrt{|\dot{f}(t)|}.$ For any time warping of $f$ by $\gamma \in \Gamma$, the SRVF of the warped function is given by $(q \circ \gamma)\sqrt{\dot{\gamma}},$ which we will denote by $q \star \gamma$ for convenience. Then, the warping functions needed to align a collection of curves $f_1,\ldots, f_n$ with corresponding SRVF representations, $q_1,\ldots,q_n$, to the Karcher mean $\mu_q^{\ast}$ are computed by solving the optimization problem:
\[
\mu_q^{\ast} = \argmin_{q\in\mathbb{L}^2}\sum_{i=1}^n \left(\inf_{\gamma\in\Gamma}\|q-q_i\star\gamma_i\|^2\right).
\]

The output of this optimization problem is the Karcher mean $\mu_q^{\ast}$ and the set of optimal warping functions $\{\gamma_i^{\ast}\}$. The optimization the problem is solved using \textit{dynamic programming}: the time domain \([0,1]\) is discretized into \(N\) equally spaced bins, and each warping function \(\gamma_i\) is approximated by a piecewise linear mapping from \((0,0)\) to \((1,1)\). This discretization corresponds to a path through an \(N \times N\) grid, where each step is required to have a positive slope to ensure invertibility.

In our analysis, we found substantial benefit in penalizing the amount of elasticity in the alignment of the curve. This is achieved by modifying the optimization problem to include a penalty term on the roughness of the $\gamma_i$ as follows:
\[
\mu_q^{\ast} = \argmin_{q\in\mathbb{L}^2}\sum_{i=1}^n \left(\inf_{\gamma\in\Gamma}\|q-q_i\star\gamma_i\|^2+ \lambda\mathcal{R}(\gamma_i)\right)
\]

The penalty $\mathcal{R}(\gamma_i)$, controlled by the constant $\lambda >0$, imposes a constraint on $\gamma_i$. In our approach, we restrict the second derivative of $\gamma_i$, so that $\mathcal{R}(\gamma_i) = \int_0^1 \ddot{\gamma_i}(t)\,dt$. This places a restriction on the smoothness in $\gamma_i$ and has the effect of keeping $\gamma_i$ closer to the identity warp $\gamma_i(t) = t$, thus regulating the level of elasticity in the alignment. The case $\lambda = 0$ is referred to as \textit{fully elastic alignment}, while $\lambda = \infty$ is the non-elastic case. 

The penalty is computed along the grid path and incorporated into the total cost, which is then minimized using dynamic programming in the same manner as without the penalty. This approach is similar to the method described by \citet{wu.2011.information}, where further details on the algorithm are provided.

An interesting alternative to adding a smoothness penalty is a metric learning approach that considers the broader 1-parameter family of elastic metrics, which extend SRVFs as described in \citet{bauer.2024.elastic}. This family allows for flexible control over warping by varying the transformation in the metric, thereby implicitly enforcing smoothness without requiring a penalty term on the $\gamma_i$.

For an intuitive overview of the elastic shape analysis procedure and a more detailed derivation of the Karcher mean, see Section~{9.1.3} of \citet{ooda}. A thorough comparison of functional data analysis with and without phase-amplitude separation is provided in Chapters 2.1, 5.4, and 9 of \citet{ooda}, demonstrating its importance for data exhibiting both types of variation. Given these established results, the observed phase variation in the GRF data motivated our application of elastic shape analysis, from which we obtained a decomposition into amplitude and phase data objects. The following subsection details our implementation and selection of the penalty parameter $\lambda$.

\subsubsection{Implementation and penalty parameter selection} The elastic shape analysis procedure was implemented via the \textsc{fdasrsf} Python package \citep{tucker-wu-srivastava:2013}. One possible approach to registering the GRF curve data is to apply the elastic shape alignment to each GRF component (vGRF, apGRF, mlGRF) separately. An analysis of amplitude and phase using this component-wise registration in the IDEA study data is detailed in Section~{4.4} of \citet{siqi.dissertation}. However, that approach is less meaningful kinetically, as each component represents one direction of the same measured force. Instead, we adopted a more intuitive approach by treating the three components as a single multidimensional curve and applying elastic shape alignment to obtain a common set of warping functions. This approach allows us to focus on the phase aspects shared by all three components. Note that the Fisher-Rao mathematics extend to multi-dimensional functions \citep{srivastava2016}, where the SRVF for a vector-valued function $f(t)$ becomes:
\[q(t)=\frac{\dot{f}(t)}{\sqrt{\|\dot{f}(t)\|}}.\]

A subset of GRF curves exhibited atypical vGRF or apGRF components, which posed challenges to aligning these curves with the rest of the data. Those GRFs were atypical in the sense that the vertical component lacked the two-peak structure expected of normal gait and appeared closer to unimodal, and/or the anterior-posterior component was close to zero and relatively flat. Examples of these atypical cases are highlighted in Figure~\ref{fig:atypicalcurves}, with representative atypical cases colored by walking speed and other curves in gray. Each panel in this row corresponds to one component of the original (unaligned) GRF curves. 
\begin{figure}[H]
	\centering
        \includegraphics[width=0.99\textwidth]{plots/atypical_curves_new.pdf}
	\caption{Left to right panels: vGRF, apGRF, and mlGRF curves. Examples of atypical GRF curves are highlighted, with representative atypical cases colored by walking speed and other curves in gray. These atypical curves lack the expected two-peak vertical structure or have near-zero, flat anterior-posterior components.}
	\label{fig:atypicalcurves}
\end{figure}

To address the lack of a common underlying structure, we registered the full dataset using the penalized elastic shape analysis procedure. By adjusting the elasticity parameter $\lambda$, we aligned the full set of GRF curves without distorting the shape of the atypical curves. To determine an appropriate $\lambda$, we computed warping functions iteratively over a grid of candidate values and visually examined the alignment results, with particular attention to atypical cases. We refined the grid in regions where $\lambda$ values produced reasonable results, selecting values that avoided excessive smoothing of features in the atypical curves while still capturing meaningful phase variation. Over-alignment led to sharp corners forming a staircase-like pattern in the warping functions, indicating drastic stretching and compression of curves with differing underlying structures, such as unimodal curves.

Based on this evaluation, we selected 
$\lambda=2$ for mitigating the staircase effect without totally sacrificing alignment. To illustrate the alignment trade-off, Figure~\ref{fig:FRresults} compares results for three $\lambda$ values: the fully elastic case $\lambda=0$ in the top row, our selected $\lambda=2$ in the middle row, and a less elastic case $\lambda=4$ in the bottom row. Atypical cases are colored by speed, while other curves are shown in gray, as in Figure~\ref{fig:atypicalcurves}. The leftmost panels display the warping functions for each $\lambda$, where $\lambda=0$ results in sharp staircase-like patterns indicative of over-alignment, while $\lambda=4$ yields warping functions tightly clustered around the identity, suggesting insufficient alignment. The second through fourth columns show the aligned curves. Results for additional values of $\lambda$ can be found in Figure~{11} of the supplementary material.

While our approach treated alignment as agnostic to a specific performance criterion, a more formal selection method could be constructed depending on the analytical objective. For example, if the interest is in evaluating alignment quality in reference to a predictive or inferential task, $\lambda$ could be optimized via cross-validation to maximize a relevant performance metric. However, in finite samples, cross-validation is prone to noisy selection due to its slow convergence to optimal results (see \citet{hall.1987.noisybw} in the context of bandwidth selection for kernel density estimation). Alternatively, if the goal is to estimate an underlying common signal, \citet{kim.2023.ppd} provides a scale-space approach for estimating both the shape of the unknown signal and the signal itself. Defining a notion of optimality depends on the specific context of the analysis, and the choice of $\lambda$ for different analytical objectives is an open question for further research.

\begin{figure}[H]
	\centering
	\includegraphics[trim=2cm 4cm 2cm 4cm,clip=true,width=0.99\textwidth]{plots/adaptiveAlignment_new.pdf}
	\caption{Alignment results for three elasticity parameter values: fully elastic ($\lambda=0$), the selected $\lambda=2$, and a more rigid case ($\lambda=4$). Atypical cases are colored by speed, while other curves are shown in gray. The leftmost panels display the warping functions, and the second through fourth columns show the aligned curves. The staircasing effect in the fully elastic case ($\lambda=0$) indicates over-alignment, while $\lambda=4$ produces warping functions tightly close to  identity, suggesting insufficient alignment. The selected $\lambda=2$ balances these effects.
}
	\label{fig:FRresults}
\end{figure}

Next, we obtained amplitude objects by applying the (common) set of warping functions to each set of original GRF curves, which provides an intuitive representation of amplitude in each component. The amplitude objects obtained for different values of $\lambda$ are shown in the second through fourth columns of Figure~\ref{fig:FRresults}.

\subsection{Modes of variation}\label{ss:modes}
In OODA terminology, a collection of members of the object space that summarize one component of variation and is in some sense one-dimensional is called a \textit{mode of variation}. For example, in the vector matrix case, a mode of variation is a rank-one matrix. We can obtain modes of variation through \textit{Principal Component Analysis} (PCA), where each object is considered as a point in high dimensional space (column vector). For an introduction to PCA, see \citet{jolliffe.2002.springer}. Amplitude modes of variation \citep{tucker-wu-srivastava:2013} were obtained for each direction of the ground reaction force (vGRF, apGRF, mlGRF) computing PCA on the set of 2,686 Fisher-Rao aligned curves, each corresponding to an individual gait observation. The sets of input curves are shown in the second, third, and fourth columns of the middle row of Figure~\ref{fig:FRresults}.

\begin{figure}[H]
	\centering
	\includegraphics[trim=0cm 0cm 0cm 0cm,clip=true,width=0.85\textwidth]{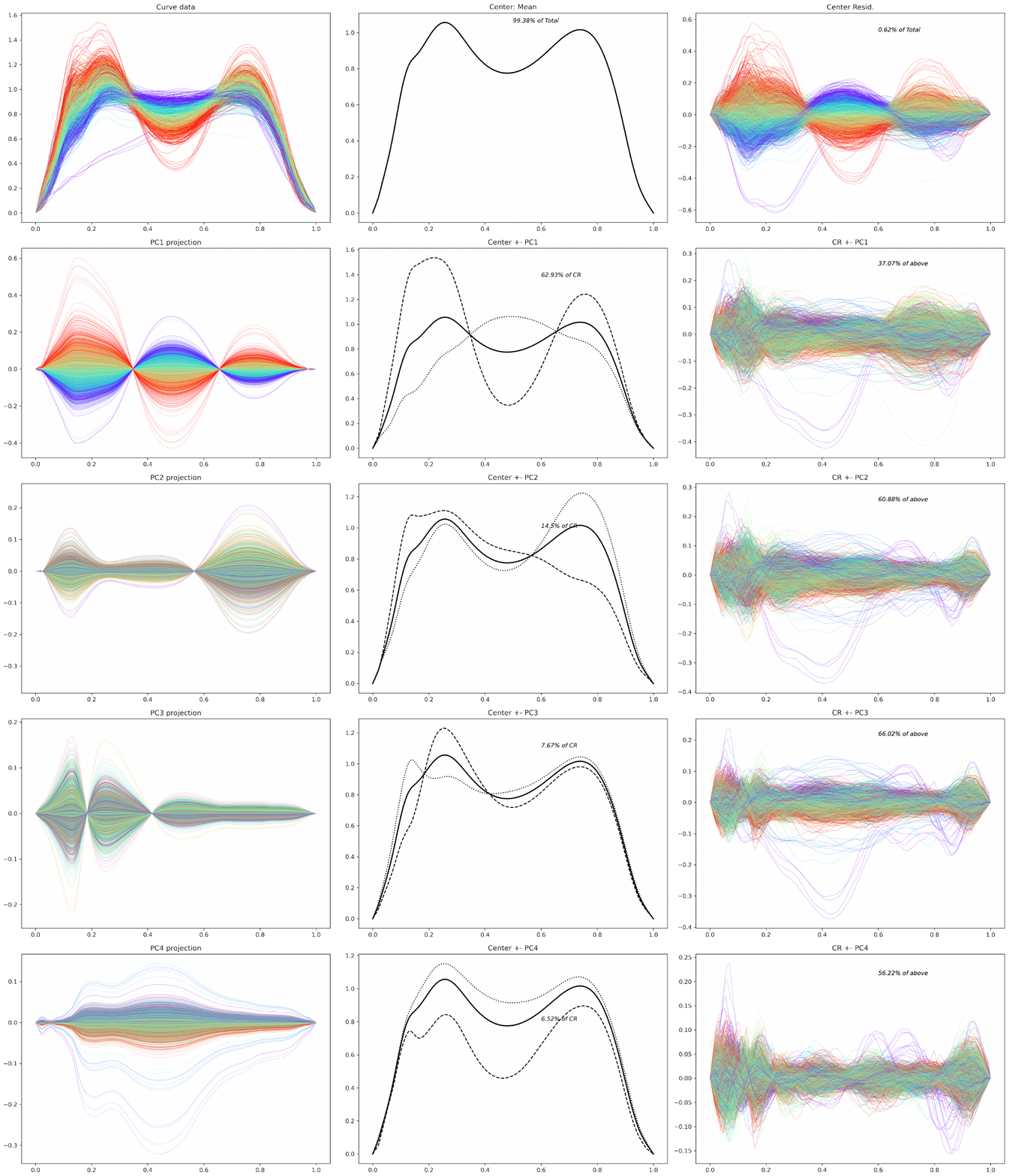}
	\caption{Modes of variation of the vGRF amplitude objects, shown in the top left panel. The first mode of amplitude variation, showing magnitude of the peaks and valley, is displayed in the second row, first panel. The panels in the middle column show the largest and smallest PC projections added back to the mean curve, while the solid black curve corresponds to the mean curve. The second and third modes of amplitude variation (third and fourth row, respectively) explain variation within each peak. The fourth mode indicates variation in overall magnitude, particularly of the valley, as seen in the middle panel of the last row.}
	\label{fig:Vamplitude}
\end{figure}

Figure~\ref{fig:Vamplitude} shows the modes of variation of the amplitude objects of the vGRF, where the curves are colored according to walking speed. The first mode of variation, shown in the first panel of the second row of that figure, is associated with walking speed and reflects the contrast in peak heights and valley depths. Faster walkers (indicated in red in the rainbow color scheme) generally exhibit higher peaks and lower valleys, while slower walkers (purple in the rainbow color scheme) have lower peaks and a shallower valley. The middle column of that figure displays the largest (dashed curve) and smallest (dotted curve) PC projections added back to the mean curve, which is shown as a solid black curve in each of the middle panels. The middle panel of the first mode shows that the slowest walkers (dotted curve) exhibit a vertical amplitude force that appears unimodal and does not exceed body weight (1 on the vertical axis), indicating that these walkers do not fully transfer their weight to the striking limb. This type of gait can be thought of as ``shuffling." The second and third modes of variation are about the second and first peak, respectively. The middle panel in the third row shows the largest and smallest PC2 projections added back to the mean curve, distinguished with a dashed (largest) and dotted (smallest) line type. These extremes show that variation in this mode is mostly in the height of the second peak. In the panel below, the extremes of the third mode indicate phase variation in the first peak that is unique to the vertical component. The fourth mode of variation reflects the overall magnitude, particularly in the mid-stance phase. The second, third, and fourth PC projection extreme curves all suggest that some curves have a small third bump before the first peak. In gait analysis, this pattern is known as the \textit{heel-strike transient} (HST), a rapid and transient rise in the vGRF immediately after ground contact. As discussed in \citet{blackburn.2016.hst} and references therein, the presence and characterization (e.g. magnitude) of HST can indicate impulsive loading, which influences cartilage degradation and symptoms of OA. However, \citet{blackburn.2016.hst} also noted that methods for identifying HST can be unreliable. The amplitude modes of variation we identified offer a potentially viable method for reliably identifying and understanding the HST.

Similar plots of the amplitude modes of variation of the components of apGRF and mlGRF are provided in Section~{2} in the supplementary material. 

Extracting phase modes of variation requires more careful consideration. Recall from Section~\ref{ss:registration} that the warping functions all have corresponding SRVFs that lie on the surface of a high-dimensional sphere in the function space. Thus, using PCA to identify phase modes of variation is essentially an approximation in the tangent space centered on the Karcher mean. In the case of warping functions, these SRVFs must also lie on the positive orthant \cite{wu.2024.test}. It is demonstrated in \citet{yu.2017.fdapns} that in cases of high variation, this tangent plane PCA may yield a distorted analysis, resulting in modes of variation that leave the positive orthant and consequently produce invalid warping functions. This phenomenon was also observed in our dataset. In such scenarios, a better decomposition of the variation can be achieved using the functional PCA methodology proposed by \citet{yu.2017.fdapns}, which is based on an improved PCA analogue for spheres known as Principal Nested Spheres (PNS) proposed by \citet{jung.2012.biometrika}. The PNS decomposition sequentially provides the best $k$-dimensional approximation $U^k$ of the data for all $k=d-1, d-2,\ldots,0$ such that 
\[
S^d \supset U_{d-1} \supset \ldots \supset U_{1} \supset U_0.
\]
For each $k$, the sphere $U^k$, called the $k$-dimensional \textit{principal nested sphere}, is a submanifold of the higher dimensional principal nested spheres. The algorithm to find sample principal nested spheres is determined by iteratively minimizing an objective function to find the best-fitting subsphere, projecting the data to the lower dimensional sphere, and mapping to the original space through a relevant transformation. The signed residuals, defined as the signed length of the minimal geodesic joining the (projected and transformed) data points to the subsphere, serve as analogs of principal component scores. Chapter 8 of \cite{ooda} provides further review of PNS and other geodesic-based methods.

\begin{figure}[H]
	\centering
	\resizebox{\textwidth}{!}{\includegraphics{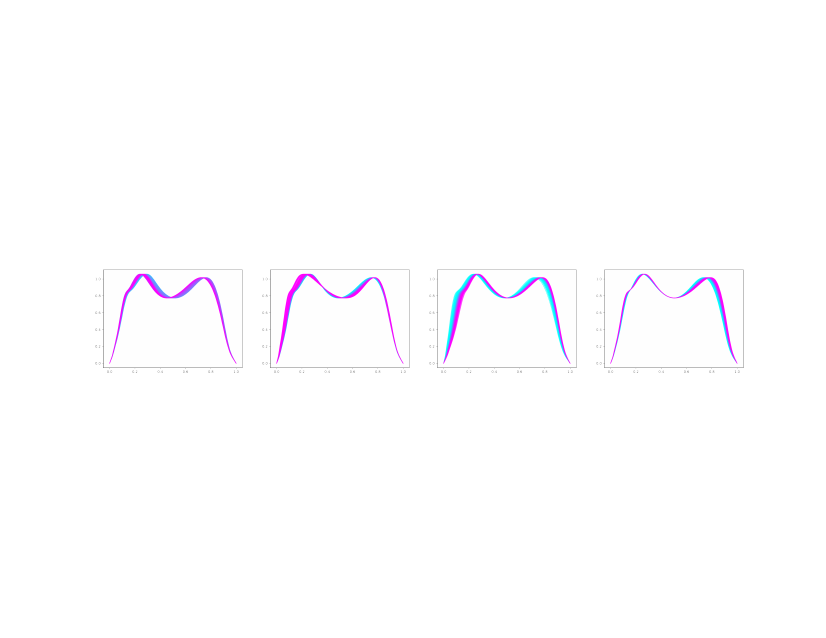}}
	\caption{Visual representation of phase variation using warpings of the Karcher mean of the vGRF curves. Warping functions were generated by taking the inverse of the phase PNS projections added to the 45-degree line. Each panel shows curves colored by corresponding PNS scores, with cyan indicating the lowest scores and magenta indicating the highest. Curves are plotted in the order of the corresponding score to avoid over-plotting. The first panel (first mode) shows an overall shift in timing, with most apparent differences in the timing of the first peak (maximum heel-strike force) and valley. The second panel (second mode) shows variability in the closeness of the peaks. The third panel (third mode) represents an overall phase shift, suggesting earlier timing for curves with a small third bump before the first peak (cyan). The fourth panel (fourth mode) emphasizes variability in the timing of the second peak, independent of the rest of the curve.}
	\label{fig:phasePNSmodes}
\end{figure}

We applied the PNS-based functional PCA methodology to the set of (common) warping functions to obtain phase modes of variation. We found that the great sphere decomposition from PNS yielded the most interpretable phase modes of variation because of weak interpretability of small sphere variation. Figure~\ref{fig:phasePNSmodes} depicts an intuitively useful notion of phase variation represented by warpings of the Karcher mean of the vGRF curves. The warping functions used to create these visualizations were generated by taking the inverse of the phase PNS projections added to the 45-degree line (identity warp). In each panel of the figure, the curves are colored based on the PNS scores for the corresponding mode, with cyan indicating the lowest scores and magenta indicating the highest. It is important to note that the curves are plotted in the order of the corresponding score, as over-plotting is an issue. The first mode (first panel) shows an overall shift in timing, with most apparent differences in the timing of the first peak (maximum heel-strike force) and valley. The second mode in the next panel appears to explain variability in the closeness of the peaks: the cyan curves are the curves with peaks closer together and the magenta curves have peaks farther apart. The third mode represents an overall phase shift (left vs. right) and seems to suggest that curves having a small third bump before the first peak correspond with earlier timing (cyan curves), especially an earlier second peak. The fourth mode appears to highlight variability in the timing of the second peak, independent of the rest of the curve.

\section{Comparison to Conventional Discrete Summaries}\label{sec:compare}
We investigated the added value of analyzing patterns across the entire movement curve, rather than relying on conventional discrete summaries of GRF curves. For this purpose, we compared how strongly full-curve modes versus conventional discrete summaries were associated with OA-related clinical traits using a nested regression framework, detailed in Section~\ref{ss:anova}. While some of the traits we considered are clinically meaningful to predict from gait biomechanics, the aim of this analysis was to quantify and compare the strength of associations between OA-related traits and gait features derived from full-curve analysis versus conventional discrete summaries. In this section, we outline the sets of gait features that served as independent variables in our models, and in the following section, we introduce the OA-related clinical traits.

To create independent variables derived from our full-curve analysis, we combined scores for $16$ distinct modes of amplitude variation from three types of curve data objects (vGRFs, apGRFs, and mlGRFs) and phase variation. Each mode is represented by either a set of amplitude PC scores or PNS phase scores. Below are listed the $16$ sets of scores that together formed our full-curve independent variables:

\begin{itemize}
    \item PC1-PC4 scores of the vGRFs amplitude data objects (studied in Figure~\ref{fig:Vamplitude});
    \item PC1-PC4 scores of the apGRFs amplitude data objects (studied in Figure~{12} in the supplementary material); 
\item PC1-PC4 scores of the mlGRFs amplitude data objects (studied in Figure~{13} in the supplementary material);
    \item Great sphere PNS1-PNS4 scores of the common phase data object (studied in Figure~\ref{fig:phasePNSmodes}).
\end{itemize}
    
We developed a set of independent variables based on conventional discrete summaries of GRF curves found in the literature, including the first peak of the vGRF curve (maximum over $0-50\%$ of the stance), the minimum value of the apGRF curve (minimum over $0-100\%$ of the stance) and the maximum value of the apGRF curve (maximum over $0-100\%$ of stance).

\subsection{OA Clinical Traits}\label{ss:traits}

The IDEA study defined several OA disease outcomes and symptoms of interest \citep{messier.2009.bmc,messier.2013.jama}. These included mechanistic outcomes: knee joint compressive force and inflammatory biomarkers (interleukin-6 [IL-6] and C-reactive protein [CRP]); and clinical outcomes: self-reported pain and function, mobility, and health-related quality of life. Increased knee joint compressive force is known to contribute to cartilage stress and degeneration and has been associated with patterns in gait biomechanics \citep{dlima.2012.compress}. Elevated IL-6 and CRP levels are linked to chronic inflammation and have been associated with knee OA \citep{messier.2009.bmc}. Pain and function were measured using the Western Ontario and McMaster Universities Arthritis Index (WOMAC) \citep{alexandersen.2014.oac}. Mobility was assessed using walking speed and distance walked in a 6-minute trial, while health-related quality of life was evaluated using the SF-36 Physical and Mental Component Scales. These IDEA study outcomes were included as dependent variables in our regression analyses.

Although not specified as outcomes in the IDEA study, we analyzed additional biomedical measures that have been examined in other OA research. Notably, radiographic OA severity is evaluated using joint space width (JSW) and Kellgren–Lawrence grade (KLG), and both are key metrics for diagnosing and monitoring disease progression. Prior studies have reported associations between gait features and variability in these structural measures (e.g. \cite{kwon.2019.identifying,jansen.2024.sradio}), making them particularly relevant to this analysis. OA is also associated with elevated fall risk, and existing studies have examined fall risk as an outcome in patients with knee OA (for example, \cite{rosadi.2022.factors}). Furthermore, anthropometric characteristics (e.g., body weight) and sociodemographic traits (e.g., age, sex, race) are important factors in OA risk, symptoms, and treatment (see, for example, \cite{sims.2009.jwomen,allen.2022.epidemiology,chang.2024.sex}). These additional measures, together with the IDEA study outcomes, are summarized in Table~{1} of the supplementary material and are collectively referred to as \textit{clinical traits} throughout the paper. For limb-level measurements (such as joint compressive force, KLG, and JSW), we used the patient-level clinical aggregates of these measures reported in the IDEA study for our analysis \citep{messier.2009.bmc,messier.2013.jama}.  

Since the IDEA study was an 18-month clinical trial of interventions in patients with advanced knee OA,  traits were collected at different intervals throughout the study. For the purpose of this analysis, focusing on baseline values provides the most consistent basis for comparison and may better reflect OA variability in the dataset. For example, although change in JSW is a key marker of progression, prior studies suggest that knees at similar OA stages progress at comparable rates \citep{benichou.2010.change}, making baseline JSW a more meaningful indicator of disease severity for the IDEA participants. Additionally, using baseline values avoids the need to account for treatment effects, which are not of interest in this analysis.

Figure~{15} in the supplementary material shows a heat map of missing values in the baseline data, with blue lines indicating missing entries for each trait; note that JSW has a relatively high rate of missing data. Baseline JSW measurements were missing for 126 of the 454 IDEA participants. Although the reasons for these missing values were not reported in the study, comparison of the other clinical traits between participants with and without baseline JSW showed no apparent sampling bias. This left 328 participants with JSW data for our analysis. Due to the large proportion of missing values, analysis involving JSW was restricted to the complete subset, with gait variables subset accordingly. Missing values for other traits, which were relatively few, were imputed using the mean of each trait (or the mean rounded to the nearest integer for integer-valued data). Importantly, there were no missing values for traits that may be particularly sensitive to imputation, such as  categorical traits like sex and race, and low-resolution ordinal measures like KLG. The treatment of different types of dependent variables (count, binary, ordinal, continuous) for modeling purposes is discussed in detail in the following section.

\subsection{Nested Model Comparison}\label{ss:anova}
We assessed the added value of full-curve gait analysis over conventional discrete summaries by comparing nested regression models. Specifically, we evaluated whether full-curve models provided additional explanatory value beyond conventional discrete summaries using a bootstrap-based likelihood ratio test (LRT) approach. Since we treat curves as data objects, all 2,686 gait curves were used as individual observations, except when modeling JSW, which was analyzed on the subset described earlier. Each trait was treated as a dependent variable and a full model fit on the combined set of gait features (full-curve modes plus conventional discrete summaries), while reduced models were fit using either the full-curve modes alone or the conventional discrete summaries alone.

The choice of regression model depended on the type of dependent variable: count, binary, ordinal, or continuous.  Number of falls was the only count-valued trait and was modeled using Poisson regression. Sex and race were the only binary-valued traits and were fit using logistic regression. Ordinal logistic regression was used for fall-related traits, WOMAC pain and function, health-related quality of life measures, and KLG. KLG is a radiographic score of OA severity, while the other ordinal traits represent scores from Likert-type scales. All remaining traits were continuous and fit using linear regression. As detailed in the previous section, missing trait values (except for JSW) were imputed using the trait mean. For integer-valued traits, the mean was rounded to the nearest integer, whereas for continuous traits it was used directly.

To compare nested models, we computed likelihood ratio tests for two cases: (1) a full model containing both full-curve modes and conventional discrete summaries versus a reduced model with only the full-curve modes, and (2) the same full model versus a reduced model with only the conventional discrete summaries. The null hypothesis in each case is that the reduced model explains the data as well as the full model, while the alternative is that the additional independent variables in the full model significantly improve model fit. Failing to reject the null when the reduced model contains the full-curve modes suggests that discrete summaries add no explanatory value beyond the full-curve modes. Conversely, rejecting the null when the reduced model contains the discrete summaries is evidence that full-curve modes provide additional explanatory value not captured by discrete summaries. The same logic applies when the roles of the two variable sets are reversed. Note that for linear regression models, the likelihood ratio test is equivalent to the nested-model F-test.

Typically, a LRT statistic is compared to a chi-squared reference distribution, but this relies on an asymptotic result which assumes independent observations. This assumption is violated in our dataset due to a clear dependence between the multiple gait curves per patient. To account for this, we implemented a bootstrap procedure in which patients were resampled with replacement, and all gait curves associated with a selected patient were included in the resampled dataset. For each of 1,000 bootstrap resamples, the regression models were on the selected sample and the corresponding LRT statistic was computed. This resampling approach reflects the natural dependence structure in the data and generates an appropriate reference distribution for inference. All other bootstrap analyses in this paper also used 1,000 replications. The significance of the observed LRT statistic from the original data was then assessed by computing the proportion of bootstrap LRT statistics greater than or equal to the observed LRT statistic. When no bootstrap replications were larger than the original statistic, we defined the p-value to be $ \frac{1}{2*1000} = 0.0005.$

 \begin{figure}[H]
	\centering
	\includegraphics[width=0.65\textwidth]{plots/plotNestedTests.pdf}
	\caption{Scatter plot of bootstrap p-values resulting from LRTs of nested models, shown on a logarithmic scale. The horizontal axis represents the p-values for testing full-curve reduced models, and the vertical axis represents those for testing conventional discrete summary reduced models. For readability, the axes are labeled on the original scale. Gray circles represent traits where neither reduced model was rejected at this significance level, indicating that neither variable set adds substantial explanatory value beyond the other. Green plus signs represent traits where the full-curve reduced model was not rejected, but the conventional discrete summary reduced model was rejected, highlighting the added value of the full-curve approach. Note that the full-curve reduced model was never rejected and is consistently a suitable approach.}
	\label{fig:nestedtestplots}
\end{figure}

Figure~\ref{fig:nestedtestplots} presents a scatterplot of p-values from the LRTs of nested models, displayed on a logarithmic scale. The horizontal axis shows p-values for testing full-curve reduced models, while the vertical axis corresponds to p-values for conventional discrete summary reduced models. For readability, the axes are labeled on the original scale. 

The green plus signs highlight traits for which the full-curve reduced model was not rejected, while the conventional discrete summary reduced model was rejected, meaning conventional discrete summaries provided no additional explanatory value. The corresponding traits include joint compressive force; both measures of radiographic OA severity (JSW and KLG); all of the anthropometric and mobility measures; and sex and race. Balance confidence, one of the fall-related traits, was also in this group, though close to the 0.05 threshold. 

The gray circles represent cases in which neither reduced model was rejected, which account for roughly half of the traits. These include both inflammatory biomarkers; self-reported pain and function; health-related quality of life measures; all but one fall-related trait; and age. Except for the biomarkers and age, these traits are Likert-type scale scores, which are highly subjective and generally difficult to model. Although inflammatory biomarkers and age are important in OA, there is no known direct relationship with gait biomechanics. Therefore, it is unsurprising that more complex gait variables do not provide significant explanatory value beyond simple discrete summaries for the gray-circle traits.

No points fall to the left of the vertical red dashed line, meaning there were no instances in which the full-curve reduced model was rejected while the conventional discrete summary reduced model was not. Although sex (Male) is near the 0.05 threshold for the full-curve reduced model, it is well below the threshold for the conventional discrete summary reduced model. Overall, these results show that a model based solely on full-curve modes is a consistently suitable approach.

\section{Conclusion}
This paper quantitatively demonstrates the extent to which complete GRF curves, compared with conventional discrete summaries, capture information relevant to disease severity and clinical profiles of OA, demonstrating the added value of full-curve analysis. We apply a straightforward nested model comparison to highlight this difference. Furthermore, our shape-based approach illustrates an intuitive representation of full movement curves that is applicable in broader analyses and reveals insightful modes of variation. To our knowledge, this work is among the first to show that analysis of full movement curves yields stronger associations with OA outcomes and OA-related clinical traits than conventional discrete summaries.

\bigskip
\begin{center}
{\large\bf SUPPLEMENTARY MATERIAL}
\end{center}

\begin{description}

\item[Supplement to Elastic Shape Analysis of Movement Data:] Additional figures and tables, with accompanying brief discussions. (pdf)

\end{description}

\bigskip
 \begin{center}
 {\large\bf Funding}
 \end{center}
 The research reported in this paper was partially supported by the National Institutes of Health (NIH), including grants from the National Institute of Arthritis and Musculoskeletal and Skin Diseases (NIAMS) P30 AR072580 and K24 AR081368. Ashley N. Buck's research was partially supported by NIAMS T32 AR082310. Addtionally, the research of Jan Hannig and J. S. Marron was also partially supported by the National Science Foundation grant DMS-2113404. The IDEA study, from which the data presented in this paper was obtained, was supported in part by NIH grants R01 AR052528-01 from NIAMS, P30 AG21332 from the National Institute on Aging, and M01-RR00211 from the National Center for Research Resources, as well as by General Nutrition Centers.

\bigskip
 \begin{center}
 {\large\bf Conflict of Interest Statement}
 \end{center}
The authors report there are no competing interests to declare.

\bigskip
 \begin{center}
 {\large\bf Statement on Generative AI}
 \end{center}
ChatGPT 4.0 was used for minor coding assistance.   
 
\bibliography{bib}
\end{document}